\documentclass[conference]{IEEEtran}
\IEEEoverridecommandlockouts
\usepackage{cite}
\usepackage{amsmath,amssymb,amsfonts}
\usepackage{graphicx}
\usepackage{textcomp}
\usepackage{xcolor}
\usepackage{url} 
\usepackage{booktabs} 
\usepackage{tabularx} 

\def\BibTeX{{\rm B\kern-.05em{\sc i\kern-.025em b}\kern-.08em
    T\kern-.1667em\lower.7ex\hbox{E}\kern-.125emX}}
\begin{document}

\title{Demo: ViolentUTF as An Accessible Platform for Generative AI Red Teaming\\
}

\author{\IEEEauthorblockN{Tam n. Nguyen}
\IEEEauthorblockA{
\textit{CISSP, IEEE member, US Federal Employee }\footnote{text for footnote}\\
tom.nguyen@ieee.org}\\
\footnotesize \textsuperscript{*}This work does not represent the views of the US Government.}

\maketitle

\begin{abstract}
The rapid integration of Generative AI (GenAI) into various applications necessitates robust risk management strategies which includes Red Teaming (RT) - an evaluation method for simulating adversarial attacks. Unfortunately, RT for GenAI is often hindered by technical complexity, lack of user-friendly interfaces, and inadequate reporting features. This paper introduces Violent UTF - an accessible, modular, and scalable platform for GenAI red teaming. Through intuitive interfaces (Web GUI, CLI, API, MCP) powered by LLMs and for LLMs, Violent UTF aims to empower non-technical domain experts and students alongside technical experts, facilitate comprehensive security evaluation by unifying capabilities from RT frameworks like Microsoft PyRIT, Nvidia Garak and its own specialized evaluators. ViolentUTF is being used for evaluating the robustness of a flagship LLM-based product in a large US Government department. It also demonstrates effectiveness in evaluating LLMs' cross-domain reasoning capability between cybersecurity and behavioral psychology.
\end{abstract}


\section{Introduction}
Generative Artificial Intelligence (GenAI) offers transformative potential across numerous domains, but its deployment is accompanied by significant risks\cite{b1}. These models can inadvertently generate harmful, biased, or inappropriate content, leak sensitive information, or be manipulated through adversarial attacks like prompt injection and jailbreaking\cite{b2}. Ensuring the safe and responsible deployment of GenAI requires rigorous evaluations and tests to proactively identify and mitigate these vulnerabilities \cite{b3}. When applied to GenAI, Red Teaming (RT) systematically probes models to elicit undesirable behaviors, assess their robustness against manipulation, and verify compliance with safety and ethical guidelines\cite{b4,b5}. However, the current landscape of tools for GenAI RT presents significant barriers\cite{b6,b7}. Table \ref{tab:ai_tools} provides a brief overview of some tool categories with their pros and cons.\cite{b8}:

\begin{figure}[htbp] 
    \centering
    \includegraphics[width=0.7\linewidth]{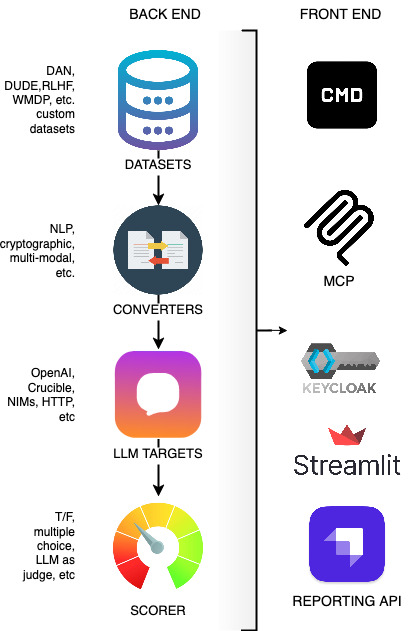} 
    \caption{Violent UTF System Components.}
    \label{fig:system}
\end{figure}

\begin{table*}[htbp]
\caption{Examples of AI Safety, Security \& Red Teaming Tools}
\label{tab:ai_tools}
\begin{center}
\begin{tabularx}{\textwidth}{>{\raggedright\arraybackslash}p{3.5cm} >{\raggedright\arraybackslash}p{5cm} >{\raggedright\arraybackslash}p{4cm} >{\raggedright\arraybackslash}X}
\toprule
\textbf{Tool/Category} & \textbf{Description} & \textbf{Pros} & \textbf{Cons} \\
\midrule
Microsoft PyRIT & Open-source Python framework for automated red teaming, orchestrating attacks. & Structured; extensible; integrates datasets; automates workflows. & Primarily code-based; learning curve for complex orchestration. \\
\addlinespace
Nvidia Garak \cite{b7} & Open-source framework to evaluate LLM security/robustness using probes, detectors, and harnesses. & Structured evaluation; modular components (probes, detectors); good for vulnerability scanning. & Primarily code-based; configuration can be complex. \\
\addlinespace
Adversarial Robustness Toolbox (ART) \cite{b7, b8} & Python library for evaluating/defending ML models against adversarial attacks (evasion, poisoning). & Wide range of attacks/defenses; supports multiple frameworks; research-oriented. & Requires ML expertise; focus broader than just LLMs; can be complex to configure. \\
\addlinespace
NB Defense \cite{b7} & JupyterLab extension/CLI for AI vulnerability management (secrets, PII, CVEs). & Integrates directly into developer workflow; contextual guidance. & Focused on notebook/code scanning; specific to Jupyter environment. \\
\addlinespace
LLM Input/Output Filters & Tools designed to moderate prompts sent to LLMs or filter responses from LLMs. & Can block known malicious prompts/harmful outputs; real-time protection. & Can be bypassed by novel attacks (jailbreaks); may introduce latency; risk of false positives/negatives. \\
\addlinespace
Llama Guard / Purple Llama \cite{b8} & Meta's tools for input/output moderation, insecure code detection. & Open-source; specifically designed for LLM safety; includes benchmarks. & Effectiveness depends on model updates and attack evolution. \\
\addlinespace
Vigil / Rebuff \cite{b8} & Python libraries/APIs focused on detecting prompt injections, jailbreaks. & Target specific attack vectors; some offer multiple detection layers (heuristics, LLM analysis). & Narrow in scope; effectiveness against novel attacks varies. \\
\bottomrule
\end{tabularx}
\end{center}
\end{table*}

Despite the availability of these tools, many demand deep technical expertise and programming skills, excluding non-technical domain experts and stakeholders who possess valuable insights into potential real-world harms \cite{b1}. Furthermore, the lack of intuitive interfaces and comprehensive reporting features limits collaborative efforts and hinders effective decision-making based on testing outcomes. \textbf{To bridge this gap, there is a clear need for a red teaming platform that democratizes the process, making it accessible to a wider range of users while providing powerful, flexible, and extensible capabilities.} This paper demonstrates how Violent UTF can meet these needs.

\section{Solution: The Violent UTF Platform}
Violent UTF aims to overcome the described obstacles in GenAI red teaming. It provides a unified, accessible, and comprehensive platform that integrates frameworks like Microsoft PyRIT and Nvidia Garak and extends features with its own specialized evaluation tools. Its creativity lies in bridging the gap between technical security testing and human-centric risk assessment, making sophisticated red teaming accessible while providing depth for expert users and automation.

\subsection{Key Features Driving Effectiveness}
\textbf{Democratized Accessibility.} Violent UTF tackles the accessibility problem head-on. Its primary interface is an intuitive web-based GUI (Streamlit) designed explicitly to empower non-programmers – domain experts, ethicists, compliance officers – to configure and execute complex red teaming scenarios without writing code. Simultaneously, a consistent Command Line Interface (CLI) and a comprehensive RESTful API (FastAPI/Kong) provide powerful options for technical users, automation scripts, and integration into MLOps pipelines.

\textbf{Unified \& Extensible Framework.} A core innovation is the unification of diverse red teaming and evaluation methodologies within a single platform. Violent UTF integrates:
\begin{itemize}
    \item Technical Red Teaming (PyRIT \& Garak) for identifying technical vulnerabilities like jailbreaks, prompt injection, and harmful content generation.
    \item Human-Centric Evaluation (Ollabench) \cite{b9} to assess LLM reasoning in the context of human-centric interdependent cybersecurity \cite{b11}. Ollabench is a custom module that allows Violent UTF users to evaluate risks related to the LLM's grasp of security policies in realistic contexts – a critical aspect often overlooked by purely technical tools \cite{b10, b11}.
\end{itemize}
Violent UTF standardizes core concepts (Generators, Prompts, Converters, Evaluators, Orchestrators, Memory) across these integrated tools, simplifying complex testing campaigns. Prompt templating further enhances flexibility for dynamic prompt generation and dataset generation.

\textbf{Secure, Scalable \& Maintainable Architecture.} Violent UTF is underpinned by a robust architecture designed for modern security and operational demands:
\begin{itemize}
    \item \textit{Presentation Layer:} Streamlit GUI and Python CLI.
    \item \textit{Authentication \& Authorization:} Centralized IAM via Keycloak and Kong Gateway, enforcing OIDC/OAuth2 and RBAC consistently across all interfaces. This secure foundation is crucial for managing user access and enabling trusted agentic operations.
    \item \textit{Unified API Layer:} FastAPI exposes all functionalities via a versioned, documented (OpenAPI) RESTful API, enabling seamless integration and automation. Kong Gateway acts as the policy enforcement point for security, routing, and rate limiting.
    \item \textit{Logging \& Observability:} A dedicated layer ensures comprehensive, structured logging with clear levels and guidelines, supporting debugging, security monitoring, and compliance.
\end{itemize}

\section{Demo - Evaluating Cross-Domain Reasoning in Cybersecurity}
Violent UTF is being used as a RT tool for evaluating a US Government's flagship LLM-based application. Besides the "main stream" topics of RT such as jailbreaking and prompt injection, the paper wants to demo Violent UTF capabilities of evaluating whether Large Language Models (LLMs) can reason effectively and safely across domains.

A critical challenge in cybersecurity is understanding and predicting human behavior regarding compliance with security policies, often influenced by complex psychological factors \cite{b11}. Solving this challenge is crucial for applications like advanced threat modeling (especially insider threats), designing targeted security awareness training, or building realistic agent-based simulations of socio-technical systems. In this use case, the Ollabench component within Violent UTF was configured to evaluate 21 different LLMs (including commercial and open-weight models) on their ability to reason about information security compliance/non-compliance behaviors.

\subsection{The main steps}
\begin{enumerate}
    \item \textbf{Scenario Generation:} Using Violent UTF's interface powered by OpenAI LLMs, scenarios depicting hypothetical employees with specific cognitive profiles (compliant or non-compliant attributes derived from 24 behavioral theories and 38 peer-reviewed papers) were generated.
    \item \textbf{LLM Interaction:} The selected LLMs (configured as 'Generators' in Violent UTF) were presented with these scenarios and asked a series of fixed multiple-choice questions designed to test different facets of reasoning: identifying cognitive constructs, comparing compliance levels, predicting team risk dynamics, and identifying key factors for intervention.
    \item \textbf{Evaluation:} Responses were evaluated using Ollabench's metrics within the Violent UTF framework, focusing on:
        \begin{itemize}
            \item \textit{Accuracy:} Overall correctness and categorical accuracy across the different question types \cite{b9}.
            \item \textit{Wastefulness:} Token efficiency, specifically measuring tokens used for incorrect answers \cite{b9}.
            \item \textit{Consistency:} Analyzing the reliability of reasoning patterns using Structural Equation Modeling \cite{b9}.
        \end{itemize}
\end{enumerate}

\subsection{Key Findings from the Use Case Evaluation}
The results obtained via Violent UTF highlight the platform's ability to uncover critical insights into LLM cross-domain reasoning:
\begin{itemize}
    \item \textbf{Significant Reasoning Gaps:} Even leading commercial models (Gemini 1.5 Flash, GPT-4o, Claude 3 Opus) achieved only around 51\% overall accuracy on these complex cross-domain tasks, underscoring the current limitations of LLMs in reliably understanding the interplay between human psychology and cybersecurity behaviors \cite{b9}. This demonstrates Violent UTF's effectiveness in identifying capability gaps where simpler evaluations might not.
    \item \textbf{Nuanced Performance Differences:} The platform revealed stark differences in how models performed on specific reasoning tasks. For instance, models showed varied success rates in identifying cognitive paths versus predicting team risks or target factors for intervention \cite{b9}. This granular analysis is crucial for selecting models suited for specific human-centric applications.
    \item \textbf{Efficiency Variance:} Significant differences in 'wastefulness' were observed, with some models consuming considerably more tokens when providing incorrect answers \cite{b9}. This capability within Violent UTF provides practical data for optimizing cost and resource allocation when deploying LLMs evaluated for such tasks.
    \item \textbf{Consistency Correlation:} The analysis indicated that more accurate LLMs generally exhibited more consistent reasoning patterns, suggesting Violent UTF can help assess the reliability alongside the correctness of an LLM's reasoning in this complex domain \cite{b9}.
\end{itemize}

\section{Conclusion and Future Work}
Violent UTF represents a significant step towards making Generative AI red teaming more accessible, comprehensive, and effective. By prioritizing user experience for both technical and non-technical users, unifying powerful testing frameworks (including PyRIT, Garak, and Ollabench), and building on a robust, secure, and scalable architecture, it addresses critical gaps in the current tooling landscape.

The provided use case demonstrates the practical applicability and efficacy of Violent UTF for organizations needing to evaluate LLMs for tasks requiring sophisticated, cross-domain reasoning. By integrating custom tools, the platform moves beyond standard security testing to provide theory-grounded, cross-domain testing in cybersecurity \cite{b14, b15}.

Future work will focus on expanding the library of integrated components (Generators, Prompts, Converters, Evaluators) by wrapping additional third-party libraries and developing novel techniques. Enhancing the reporting and visualization capabilities within the GUI is a key priority, providing users with more interactive ways to analyze results and generate actionable insights. Further development of the agentic capabilities will enable more sophisticated automated red teaming scenarios, potentially involving AI agents dynamically adapting attack strategies based on model responses. Continued adherence to API-first development, including contract testing and versioning, will ensure maintainability and ease of integration as the platform evolves towards potential microservice patterns. By fostering collaboration and lowering barriers to rigorous testing, Violent UTF aims to contribute significantly to the development and deployment of safer, more trustworthy Generative AI systems.

\appendix
Here is the supplementary material:

\begin{itemize}
    \item YouTube video demonstration of ViolentUTF: \url{https://youtu.be/c-UCYXq0rfY}
    \item Source code on GitHub: \url{https://github.com/Cybonto/ViolentUTF_nightly}
\end{itemize}

Please contact me for access to the private GitHub repository.

\vspace{12pt}

\end{document}